\documentclass[%
 aip,
 jap,%
 amsmath,amssymb,
 reprint,
]{revtex4-1}

\usepackage{graphicx}
\usepackage{epstopdf}
\usepackage{natbib}
\usepackage{dcolumn}
\usepackage{bm}

\begin{document}

\preprint{AIP/123-QED}

\title[Reversible mechanical and electrical properties of ripped graphene]{Reversible mechanical and electrical properties of ripped graphene}

\author{J. Henry Hinnefeld}
 \affiliation{Department of Physics, University of Illinois at Urbana-Champaign, 1110 West Green Street, Urbana, Illinois 61801, USA}
\author{Stephen T. Gill}
 \affiliation{Department of Physics, University of Illinois at Urbana-Champaign, 1110 West Green Street, Urbana, Illinois 61801, USA}
\author{Shuze Zhu}
 \affiliation{Department of Mechanical Engineering, University of Maryland, College Park, MD 20742, USA}
\author{William J. Swanson}
 \affiliation{Department of Physics, University of Illinois at Urbana-Champaign, 1110 West Green Street, Urbana, Illinois 61801, USA}
\author{Teng Li}
 \email{lit@umd.edu}
 \affiliation{Department of Mechanical Engineering, University of Maryland, College Park, MD 20742, USA}
\author{Nadya Mason}
 \email{nadya@illinois.edu}
 \affiliation{Department of Physics, University of Illinois at Urbana-Champaign, 1110 West Green Street, Urbana, Illinois 61801, USA}

\date{\today}

\begin{abstract}
We examine the mechanical properties of graphene devices stretched on flexible elastomer substrates. Using atomic force microscopy, transport measurements, and mechanics simulations, we show that micro-rips form in the graphene during the initial application of tensile strain; however subsequent applications of the same tensile strain elastically open and close the existing rips. Correspondingly, while the initial tensile strain degrades the devices' transport properties, subsequent strain-relaxation cycles affect transport only moderately, and in a largely reversible fashion, yielding robust electrical transport even after partial mechanical failure.
\end{abstract}

\maketitle

\begin{figure}[floatfix]
\centering
\vspace{0.2cm}
\includegraphics[width=\linewidth]{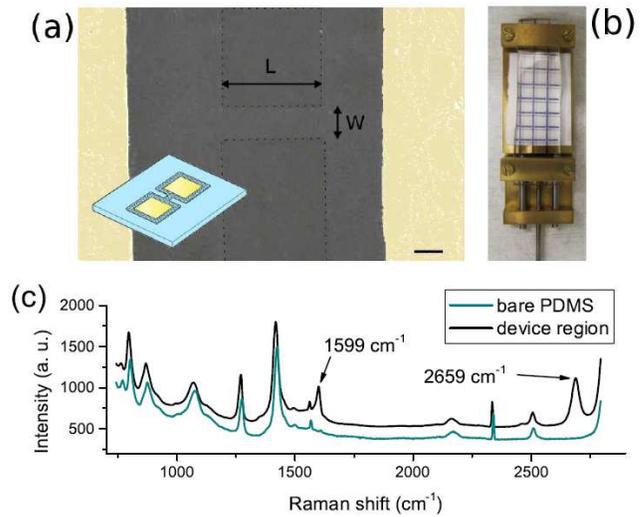}
\caption{\textbf{(a)} False-color optical image of a graphene bridge device (outlined by dashed line) patterned on a PDMS substrate with gold contact pads (light yellow). The length (L) and width (W) of the bridge are described in the text. The scale bar is 25 $\mu$m. Inset: A schematic illustration of the device geometry. \textbf{(b)}  The mechanical stretching stage with PDMS inserted between the clamps. The devices are stretched along the axis of the micro-bridge. \textbf{(c)} Offset Raman spectra for a bare PDMS region and a graphene device region. The graphene G and 2D peaks, at 1599 cm$^{-1}$ and 2659 cm$^{-1}$ respectively, in the spectra from the device region confirm the presence of graphene.}
\label{fig:device}
\end{figure}

\begin{figure*}
\centering
\includegraphics[width=\linewidth]{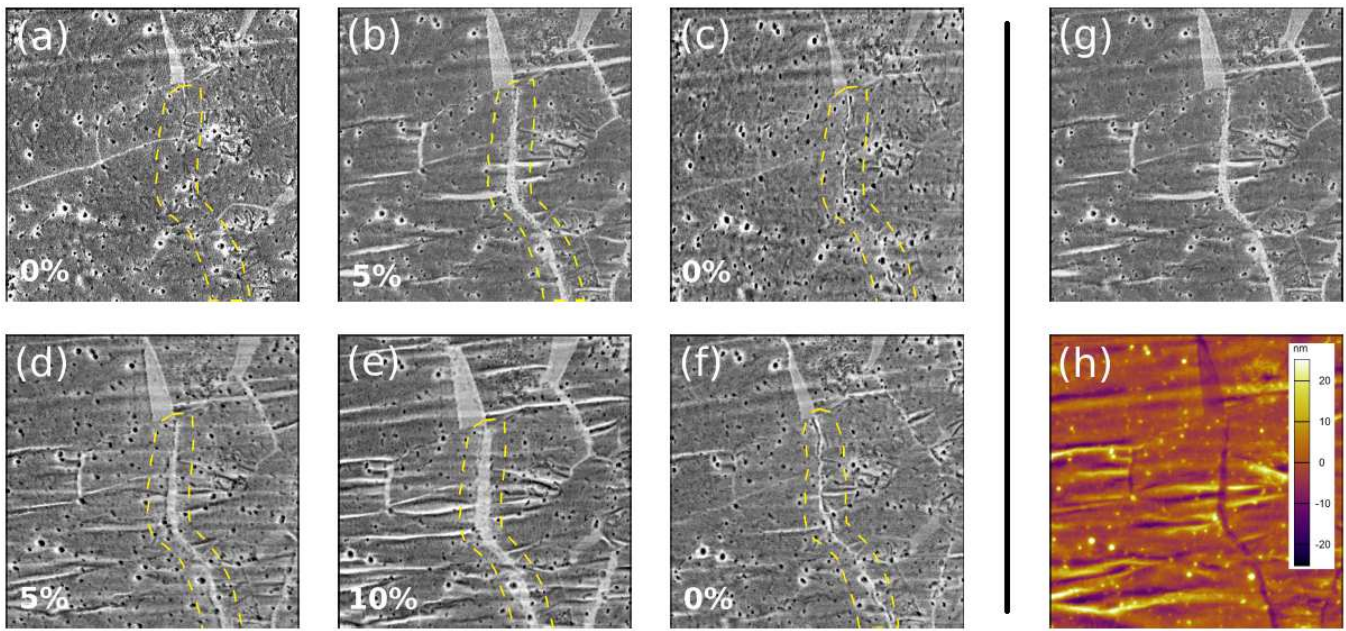}
\caption{\textbf{(a-f)} AFM phase measurements of graphene on a polymer substrate at approximately 0, 5, 0, 5, 10, and 0 percent strain (applied along the horizontal axis), as labeled. Rips are evident as light-gray, elongated vertical features. An example of a rip that opens and closes with applied strain is indicated by the dashed line. Dark spots present in each image are debris on the substrate surface; white halos surrounding some of the debris are indicative of graphene slightly delaminating from the substrate. Elongated horizontal features are strain-dependent wrinkles. \textbf{(g)} AFM phase and \textbf{(h)} height data. Variations in the height data distinguish between wrinkles and rips in the graphene, which have similar signatures in the phase data. The scanned area in each image is 25 $\mu$m$^2$.}
\label{fig:rips}
\end{figure*}

\begin{figure}
\includegraphics[width=\linewidth]{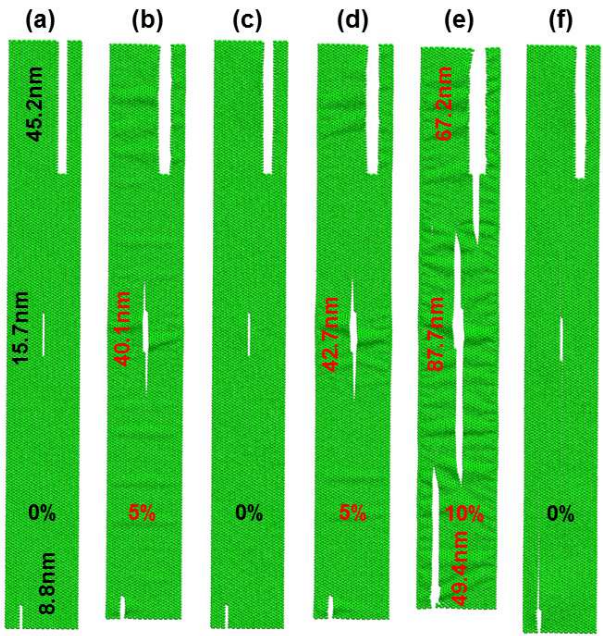}
\caption{Coarse-grained simulations show the elastic opening and closing of rips during initial and subsequent tensile loading cycles, in good agreement with AFM measurements in Figure 2. The graphene region was simulated at 0, 5, 0, 5, 10, and 0 percent strain applied along the horizontal axis, as labeled. Values given in nm refer to the rip lengths. The vertical contraction of the graphene region at higher strain values is due to the Poisson effect.}
\label{fig:Simulation}
\end{figure}

\begin{figure*}
\includegraphics[width=0.49\linewidth]{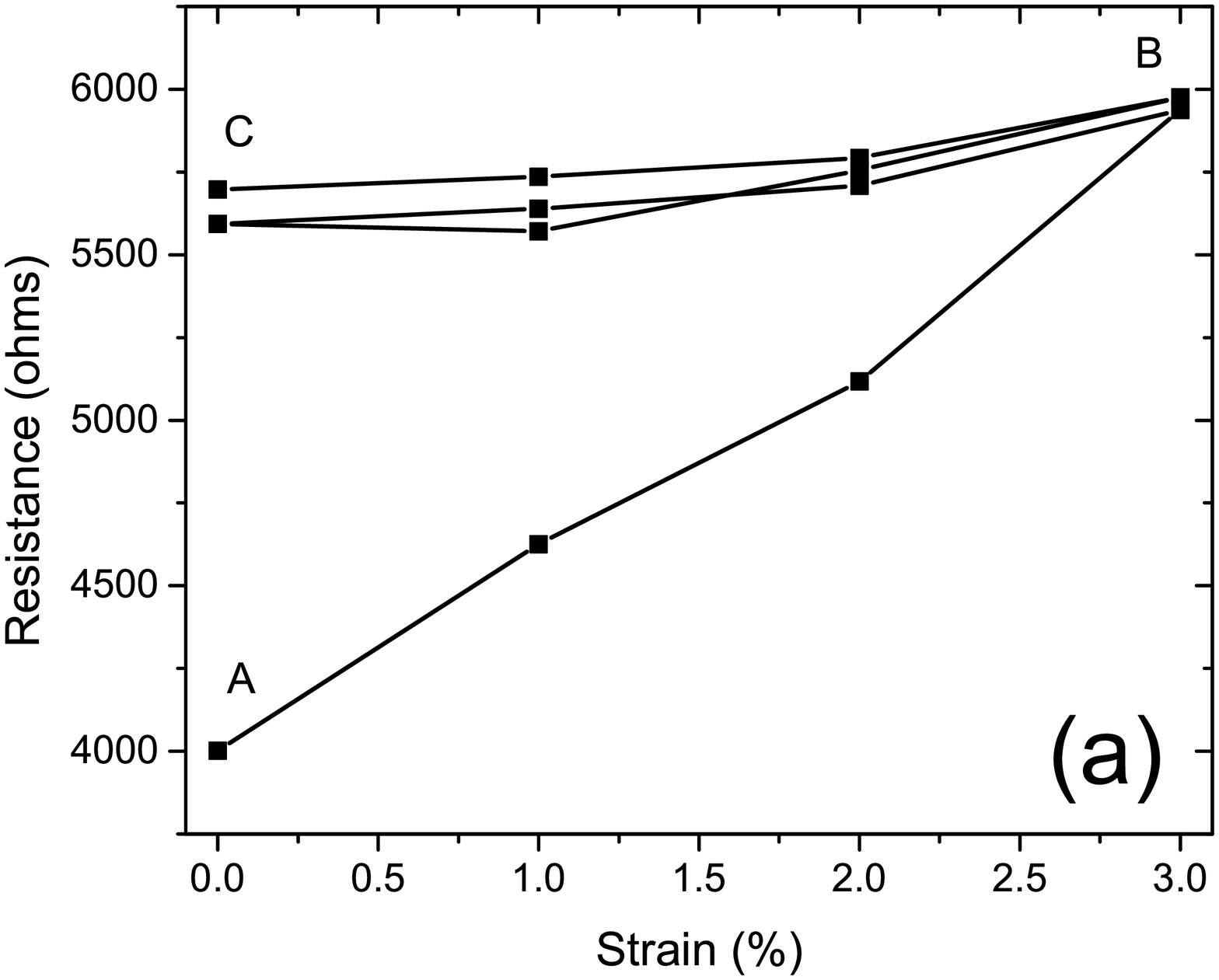}
\includegraphics[width=0.49\linewidth]{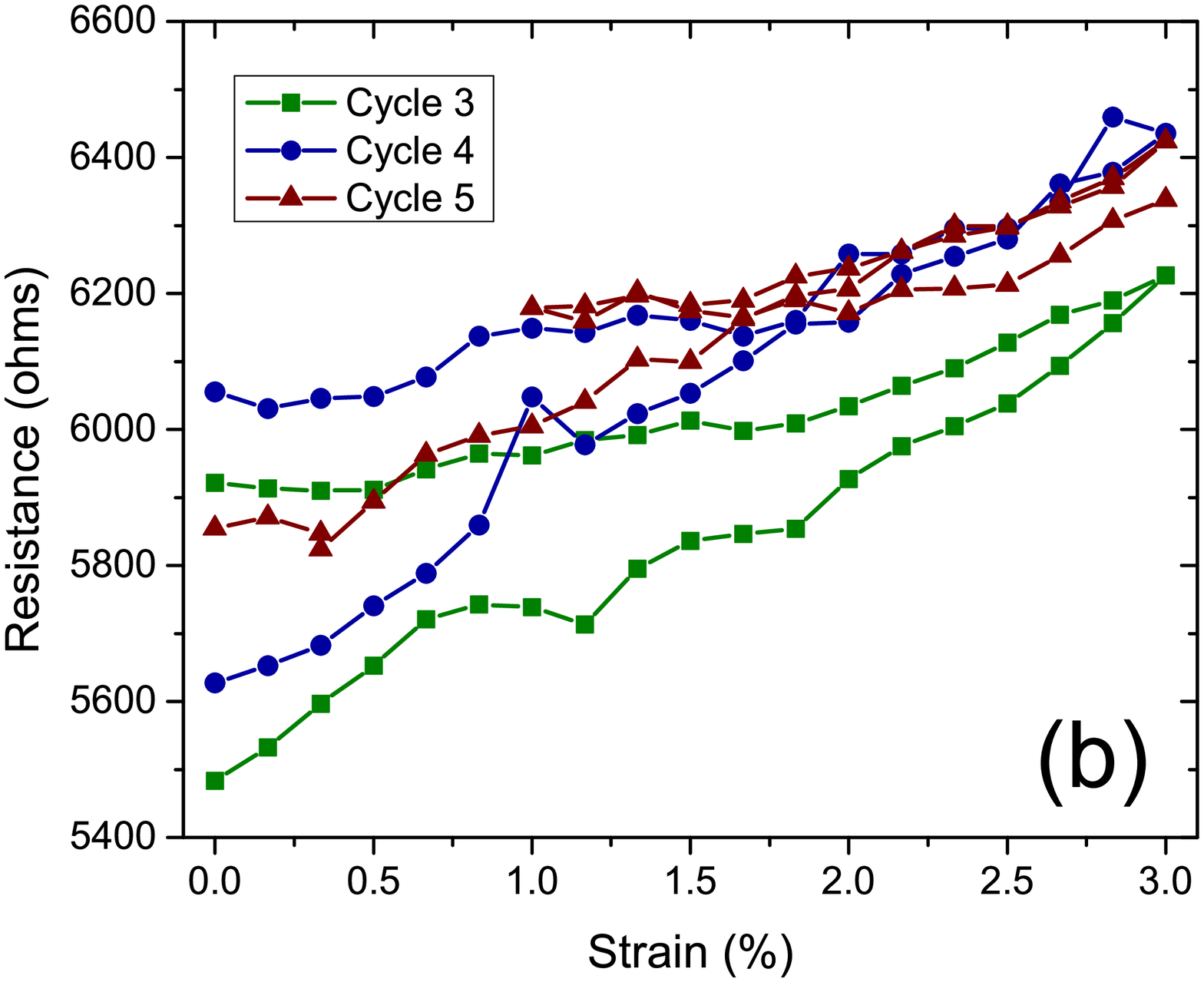}
\caption{\textbf{(a)} Electrical resistance of a graphene device vs. applied tensile strain. The initial application of strain significantly increases the resistance while subsequent strain-relaxation cycles over the same strain range yield smaller, mostly reversible changes in the resistance. \textbf{(b)} Three consecutive strain-relaxation cycles (Cycles 3,4,5), showing largely reversible transport characteristics.}
\label{fig:RvsStrain}
\end{figure*}

\section{Introduction}

Recent advances in graphene production\cite{Kim2009, Bae2010, Lee2010} have enabled the fabrication of a variety of flexible, graphene-based electronic components, including transparent interconnects\cite{Kim2011}, high-performance capacitors\cite{El-Kady2012}, and transistors\cite{Lee2011}. The prospect of flexible, graphene-based electronic devices suggested by these results raises an important question: are graphene's electrical properties and mechanical integrity robust under the strains graphene is likely to experience in such devices? Pristine graphene has an exceptionally high breaking strength\cite{Lee2008}, yet it may be susceptible to ripping, particularly if it has defects \cite{Kim2012} and/or strong  surface adhesion\cite{Sen2010}. It is still relatively unknown under what strain conditions substrate-supported graphene rips, and how the electrical properties are then altered.

In this Letter, we combine atomic force microscopy (AFM), coarse-grained mechanical simulations, and electrical transport measurements to study the effects of lateral strain on rips in graphene. We find that graphene adhered to a flexible substrate and then stretched laterally can develop small rips with only 1\% applied strain. However, even with ripping, the electrical properties remain relatively robust: introducing small rips slightly increases the resistance, but subsequent strain-relaxation cycles over the same strain range change transport only modestly, and in a largely reversible fashion. Such resilience is atypical for conducting thin films, which typically demonstrate rapid and irreversible device failure after the onset of rip formation\cite{Cairns2000,Fortunato2002}.

\section{Experimental Details}

Devices consisted of patterned graphene placed on flexible polydimethylsiloxane (PDMS) substrates. The devices were fabricated using a modified transfer printing process, similar to that described in Ref \onlinecite{Kim2009}. Single-layer graphene was grown using established chemical vapor deposition (CVD) techniques \cite{Li2009}, and then transferred to a copper-coated silicon wafer where it was patterned using photolithography and reactive ion etching. Next, a piece of PDMS was mechanically pressed onto the silicon wafer, and the copper was then etched to leave patterned graphene on the PDMS substrate\cite{Lee2010}. Raman spectroscopy was used to confirm the presence of graphene on the PDMS as shown in Figure 1c; the shape of the Raman 2D peak\cite{Ferrari2006}, as well as subsequent AFM measurements verified the single-layer character of the graphene. Finally, shadow-mask evaporation was used to deposit Ti/Au contact pads. The device geometry is illustrated in Figure 1a: a narrow graphene bridge connects two large graphene pads, each of which is covered with a Ti/Au contact pad. We studied 13 different devices having bridge aspect ratios ranging from 1.5:1 to 12:1 (length:width) and widths of 100, 50, and 25 $\mu$m. The data in this manuscript focuses on a device with a bridge width of 25 $\mu$m and an aspect ratio of 2:1. The data for all samples yielded similar qualitative results. Quantitative differences in transport data between different devices were uncorrelated with the bridge dimensions, and instead seemed to be dominated by pre-existing rips in the graphene, which are often introduced during the graphene transfer process\cite{Kim2012}.

AFM and transport measurements were performed while the PDMS substrate was mounted in a mechanical stretching stage, as shown in Figure 1b. The substrate was clamped at either end, and then strained by turning the threaded rod, which laterally moves the sliding clamp along its guide rails. A mechanical stepper motor was used to control the stretching stage position to ensure reproducibility. Variable device positioning on the substrate as well as slight variations in substrate thickness preclude exact conversion between strain applied to the substrate and to the device, therefore `turns of the stretching stage control rod' were used as the controlled variable. Each turn strains the substrate by approximately one percent, and we estimate that the strain applied to the graphene differs from that applied to the PDMS substrate by no more than ten percent. However, our conclusions are unaffected by this uncertainty, as variations in the magnitude of applied strain between devices only shift the strain axis of the data while preserving the observed trends. Optical observations indicated that the Ti/Au pad adhesion to the substrate was robust and did not slip during measurements. Transport measurements were performed by placing micro-manipulator probes in contact with the gold contact pads at each strain value, and AFM measurements were performed with an Asylum Research MFP-3D.

\section{Results and Discussion}

Figures 2a-f show AFM phase images of graphene in the bridge region of a device at 0, 5, 0, 5, 10, and 0 percent strain applied along the horizontal axis of the images. Both rips and delaminations caused by wrinkles appear as a function of strain, and can be distinguished via AFM height data: Figs. 2g and 2h show that wrinkles have corresponding undulations in the height data (peaks and dips) while rips are indicated by a uniform depression (consistent with the substrate exposed between graphene regions). In Fig. 2, the vertical features are rips and the majority of the horizontal features are wrinkles. 

The opening and closing of rips is clear in the Figure:  the unstrained device (Fig. 2a) exhibits some small rips and defects. When the substrate is mechanically stretched (Fig. 2b) the existing rips widen and new rips form; when the applied strain is relaxed (Fig. 2c), pre-existing defects return to nearly their original condition and newly formed rips close. Subsequent strain-relaxation cycles over the same strain range re-open existing rips (Fig. 2d), but proceeding to a higher strain range forms new rips and widens pre-existing ones (Fig. 2e), which then close less completely when the strain is relaxed (Fig. 2f). The strain values at which we observe micro-rip formation are substantially lower than the reported fracture strength of graphene\cite{Lee2008}, however the tensile strength of graphene is strongly susceptible to defects such as holes and tears\cite{Lee2013}. Although graphene produced by CVD is known to be polydomain, it has been shown that rips in graphene do not preferentially follow grain boundaries\cite{Kim2012}. Rather, the fabrication procedures used to generate patterned graphene devices on polymer substrates routinely introduce rips and other defects in the graphene, which accounts for the mechanical failure observed at low strain values.

To shed light on the underlying mechanism of the rip formation and evolution, we simulate rip formation and the subsequent elastic opening and closing of rips in graphene, via a coarse-grained (CG) modeling scheme\cite{Zhu2014}. Given the prohibitive simulation expense to model rips of real size in experiments (microns in length), we simulate a scaled-down model of a graphene monolayer with a size of 24 nm by 200 nm (Fig. 3). Three pre-cracks of various sizes are introduced in the model (Fig. 3a) to mimic the pre-existing defects in the as-made sample. Each CG bead in the graphene interacts with a virtual substrate via a Lennard-Jones potential\cite{Scharfenberg2011} $V_{gs}(r) =4\varepsilon_{gs} \left( \frac{\sigma_{gs}^{12}}{r^{12}} - \frac{\sigma_{gs}^6}{r^6} \right)$, where $\varepsilon_{gs}=0.01844 \, \text{eV}$ and $\sigma_{gs}=0.29 \, \text{nm}$, which gives rise to an adhesion energy around 0.044 eV/nm$^{2}$. In addition, the CG beads on the four outer edges of the simulation model are not allowed to slide relative to the substrate so that the tensile loading of the graphene can be applied by stretching the substrate along the horizontal direction, similar to the experimental setup.

As the applied tensile strain first increases to 5\%, the stress concentration near the tips of the short middle crack ($\sim$15.7 nm in length) becomes sufficiently high to cause the propagation of the short crack in both directions. Due to the nature of displacement loading, the driving force for crack propagation decreases as the crack extends. As a result, the middle crack stops advancing at a length of $\sim$40.1 nm (Fig. 3b). Upon unloading of the tensile strain the elongated middle crack closes, nearly fully recovering the original shape of the graphene (Fig. 3c); however, the atomic bond breaking in graphene during crack propagation is not reversible. Consequently, the graphene cannot fully recover its original mechanical integrity.

Further tensile loading up to 5\% causes the cracks to reopen but further extension of the cracks is shown to be negligible (Fig. 3d), largely due to a lack of sufficient driving force for crack propagation. The application of a tensile loading of 10\% provides sufficient driving force to cause all three cracks to extend significantly. The crack propagation eventually saturates due to the decreasing driving force under displacement loading (Fig. 3e). Upon unloading to zero strain, all newly formed cracks close, resulting in a graphene morphology nearly identical to its original shape (Fig. 3f), similar to the experimental observation (Fig. 2e to Fig. 2f).

Simulations also show the formation of delaminations and horizontal wrinkles in graphene upon tensile loading and the disappearance of such features upon unloading, which agrees with the experimental observations (Fig. 2). We attribute the formation of these delamination and wrinkle features to the combined effect of a mismatch in Poisson's ratios between graphene and the PDMS substrate and the relatively weak graphene/PDMS interfacial bonding. In addition, recent studies show that the location of wrinkles in graphene can be guided by the debris distribution on the substrate surface \cite{Zhu2014b}, consistent with our experimental observations in Fig. 2.

The behavior of the rips determines the electrical transport as a function of strain, as evident in Fig. 4. Figure 4a demonstrates three important features of the data: first, during the initial application of strain (A to B in the Figure) the resistance increases (for this sample, by approximately 43 percent). Typical values for this initial increase in other devices ranged from 20 to 40 percent of the starting resistance. Second, the resistance of the device decreases as the applied strain is relaxed (from B to C) by 7 percent for this device, and typically by between 6 and 14 percent. Finally, in subsequent strain-relaxation cycles over the same strain range the resistance changes only moderately, and in a largely reversible fashion.

The transport behavior can be explained by the opening and closing of rips: in the unstrained device, small rips largely determine the initial resistivity.  The device's resistance increases when the substrate is mechanically stretched, due to the widening of existing rips and formation of new ones; subsequent strain-relaxation cycles over the same strain range, which re-open and close existing rips, generate largely reversible changes in resistance. This reversibility is demonstrated in Fig. 4b; data from the same device recorded during the third, fourth, and fifth strain-relaxation cycles are shown in green, blue, and red respectively. In each case the resistance changes by $\sim$14\% for $\sim$3\% applied strain, and returns to within 8\% of its original value. Proceeding to a higher strain range forms new rips, consistent with a jump in resistance when the strain range is increased. This behavior -- an increase in resistance with the initial application of tensile strain, followed by moderate and reversible changes in the resistance during subsequent strain-relaxation cycles over the same strain region -- persists up to approximately 15\% applied strain, at which point the devices become permanently non-conducting.

Previous experimental work has demonstrated reversible transport changes in strained graphene, either by depositing graphene on pre-strained substrates so as to create controlled crumpling\cite{Zang2013} and buckling\cite{Wang2011}, by patterning complex interconnect geometries\cite{Kim2011,Lee2010}, or by measuring transport across macroscopic graphene films\cite{Kim2009,Bae2010}. In comparison, this work demonstrates the continuing robustness of device functionality \textit{after} partial mechanical failure. Such resilience is distinctly atypical for conducting thin films: similar studies performed on tin-doped indium oxide (ITO)\cite{Cairns2000} and zinc oxide\cite{Fortunato2002} reported rapid and irreversible device failure after the onset of rip formation.  One potential explanation for graphene's exceptional resilience is its morphological simplicity: as a two-dimensional membrane re-establishing electrical contact between two sides of a rip is as simple as overlaying two sheets of paper, while for typical three-dimensional thin films the process is more similar to fitting two halves of a snapped pencil back together.

\section{Conclusion}

In summary, we have observed the formation and subsequent evolution of micro-rips in graphene using atomic force microscopy. While an initial application of tensile strain introduces new mechanical defects, successive strain-relaxation cycles over the same strain range elastically open and close the existing rips. Mechanics simulations further reveal the underlying deformation and failure mechanisms of the graphene sample under initial and subsequent cyclic tensile loadings, which agree well with the AFM measurements. This mechanical effect has a corresponding electrical effect: the graphene's transport properties are degraded by the initial application of strain, but show small, mostly reversible changes during ensuing strain-relaxation cycles. Graphene's combination of superlative electronic properties and robust functionality after partial mechanical failure is unique among conducting thin films and has promising implications for future device applications.

\section*{Acknowledgements}
We thank Scott Maclaren (UIUC MRL/CMM) for technical assistance. This work was supported by NSF grants \#1069076 and \#1129826 (SZ, TL), NSF-CMMI grant \#1130364 and NSF-NEB grant \#486171 (JHH, STG, WJW, NM), and was carried out in part in the Frederick Seitz Materials Research Laboratory Central Facilities, University of Illinois.

\bibliography{main}

\end{document}